%% file: EXO-10-003_temp.tex
\pdfoutput=1

\documentclass[11pt,twoside,a4paper,cmspaper,final,collab]{cms-tdr}
\def\svnVersion{23336:23344M}\def\cmsCernNo{2010-049}\def\cmsCernDate{\today}\def\cmsMessage{Submitted to Physical Review Letters}

\begin{document}\cmsNoteHeader{EXO-10-003}

\hyphenation{env-iron-men-tal}
\hyphenation{had-ron-i-za-tion}
\hyphenation{cal-or-i-me-ter}
\hyphenation{de-vices}
\RCS$Revision: 23344 $
\RCS$HeadURL: svn+ssh://alverson@svn.cern.ch/reps/tdr2/papers/EXO-10-003/trunk/EXO-10-003.tex $
\RCS$Id: EXO-10-003.tex 23344 2010-11-26 18:10:01Z alverson $

\input{ptdr-definitions}
\providecommand{\GeVbare}{\ensuremath{\text{Ge\hspace{-.08em}V}}\xspace}

\cmsNoteHeader{EXO-10-003} 
\title{Search for Stopped Gluinos in $pp$ collisions at $\sqrt{s}$ = 7 TeV}

\address[cern]{CERN}
\address[neu]{Northeastern University}
\author[cern]{The CMS Collaboration}

\date{\today}

\abstract{The results of the first search for long-lived gluinos produced in 7 TeV $pp$ collisions at the CERN Large Hadron Collider are presented.  The search looks for evidence of long-lived particles that stop in the CMS detector and decay in the quiescent periods between beam crossings.  In a dataset with a peak instantaneous luminosity of $1 \times 10^{32}~{\rm cm}^{-2} {\rm s}^{-1}$, an integrated luminosity of 10 pb$^{-1}$, and a search interval corresponding to 62 hours of LHC operation, no significant excess above background was observed.  Limits at the 95\%~confidence level on gluino pair production over 13 orders of magnitude of gluino lifetime are set.  For a mass difference $m_{\tilde{g}}-m_{\tilde{\chi}^0_1} >100$ GeV/$c^2$, and assuming BR($\tilde{g} \rightarrow g\tilde{\chi}^0_1$) = 100\%, $m_{\tilde{g}} < 370$ GeV/$c^2$  are excluded for lifetimes from  10~$\mu$s to 1000~s.}

\hypersetup{%
pdfauthor={CMS Collaboration},%
pdftitle={Search for Stopped Gluinos in pp collisions at sqrt s = 7 TeV},%
pdfsubject={CMS},%
pdfkeywords={CMS, physics, stopped gluinos}}

\maketitle 

Many extensions of the standard model predict the existence of new heavy quasi-stable particles~\cite{Fairbairn:2006gg}.  Such particles are present in some supersymmetric models~\cite{Dimopoulos:1996vz, Baer:1998pg,Jittoh:2005pq}, ``hidden valley'' scenarios~\cite{Strassler:2006im}, and grand-unified theories (GUTs), where the new particles decay through dimension five or six operators suppressed by the GUT scale~\cite{Arvanitaki:2008hq}.  Long-lived particles are also a hallmark of split supersymmetry~\cite{ArkaniHamed:2004fb}, where the gluino ($\tilde{g}$) decay is suppressed due to the large gluino-squark mass splitting, from which the theory gets its name.  Of these possibilities, the Compact Muon Solenoid (CMS) experiment is most sensitive to models like split supersymmetry where production proceeds via the strong interaction resulting in relatively large cross sections at the Large Hadron Collider (LHC)~\cite{Dawson:1983fw,Beenakker:1996ch,Plehn:2005cq,Beenakker:2009ha}.  For this reason, we have targeted the search described in this Letter at long-lived gluinos.  Existing experimental constraints on the lifetime of such gluinos are weak~\cite{Abazov:2007ht,Arvanitaki:2005fa};  these gluinos may be stable on typical CMS experimental timescales.  Lifetimes of $\mathcal{O}$(100--1000) seconds are especially interesting in cosmology since such decays would affect the primordial light element abundances, and could resolve the present discrepancy between the measured $^6{\rm Li}$ and $^7{\rm Li}$ abundances and those predicted by conventional big-bang nucleosynthesis~\cite{Arvanitaki:2008hq,Jittoh:2008eq}.
	
	If long-lived gluinos were produced at the LHC, they would hadronize into $\tilde{g}g, \tilde{g}q\bar{q}, \tilde{g}qqq$ states, collectively known as ``R-hadrons'' some of which would be charged, while others would be neutral.  Those that were charged would lose energy via ionization as they traverse the CMS detector.  For slow R-hadrons, this energy loss would be sufficient to bring a significant fraction of the produced particles to rest inside the CMS detector volume~\cite{Arvanitaki:2005nq}.  These ``stopped" R-hadrons may decay seconds, days, or even weeks later, resulting in a jet-like energy deposit in the CMS calorimeter.  These decays will be out-of-time with respect to LHC collisions and may well occur at times when there are no collisions in CMS.  The observation of such decays, in what should be a ``quiet'' detector except for an occasional cosmic ray, would be an unambiguous discovery of new physics.  	

The CMS apparatus has an overall length of 22~m, a diameter of 15~m,
and weighs $14\,000$ tons.  The CMS coordinate system has the origin at the center of the detector.
The $z$-axis points along the direction of the counterclockwise beam, with the transverse
plane perpendicular to the beam; $\phi$ is the azimuthal
angle in radians, $\theta$ is the polar angle, and the pseudorapidity is
$\eta \equiv -\ln(\tan[\theta/2])$.  The central feature of the CMS apparatus is a superconducting solenoid of 6~m internal diameter, providing a field of 3.8~T.  Within the field volume are the silicon pixel and strip tracker, the crystal electromagnetic
calorimeter (ECAL) and the brass-scintillator hadronic calorimeter
(HCAL).  Muons are measured in gas-ionization detectors embedded in the steel return yoke.  The HCAL, when combined with the ECAL, measures jets with a
resolution $\Delta E/E \approx 100\,\%\sqrt{E\,[\GeVbare]} \oplus 5\,\%$.  The
calorimeter cells are grouped in projective towers, of granularity
$\Delta \eta \times \Delta \phi = 0.087\times0.087$ at central
rapidities.  In this analysis, jets are reconstructed using an iterative cone algorithm with $R = \sqrt{\Delta\phi^2 + \Delta\eta^2} = 0.5 $. The reconstructed jet energy $E$ is defined as the scalar
sum of the calorimeter tower energies inside the jet.
The first level (L1) of the CMS trigger system, composed of custom
hardware processors, uses information from the calorimeters and muon
detectors to select (in less than 3~$\mu$s) the most interesting
events.  The High Level Trigger (HLT) processor farm further decreases the event rate to
300~Hz before data storage.  A more detailed description of the CMS experiment
can be found elsewhere~\cite{Adolphi:2008zzk}.

The 7 TeV center-of-mass $pp$ collision data analyzed in this Letter were recorded by CMS between April and October 2010.  We divide these data into two samples:  the first corresponds to 95 hours of trigger live-time during LHC ``fills'', in which the instantaneous luminosity was $2-7 \times 10^{27}~{\rm cm}^{-2} {\rm s}^{-1}$.  We use this as a control sample to estimate the background rate.  Because these data were recorded at relatively low instantaneous luminosity, there is negligible risk that a stopped-particle signal is present in this sample. 
The second sample, in which we search for the presence of a stopped-particle signal, corresponds to 62 hours of trigger live-time during which data, corresponding to an integrated luminosity of 10 pb$^{-1}$, were recorded by CMS with a peak instantaneous luminosity of $1 \times 10^{32}~{\rm cm}^{-2} {\rm s}^{-1}$.  In producing these data, the LHC was filled with up to 312 proton bunches per beam.

 \begin{figure}[th]
 \begin{center}
      \resizebox{1.0\linewidth}{0.7\linewidth}{\includegraphics{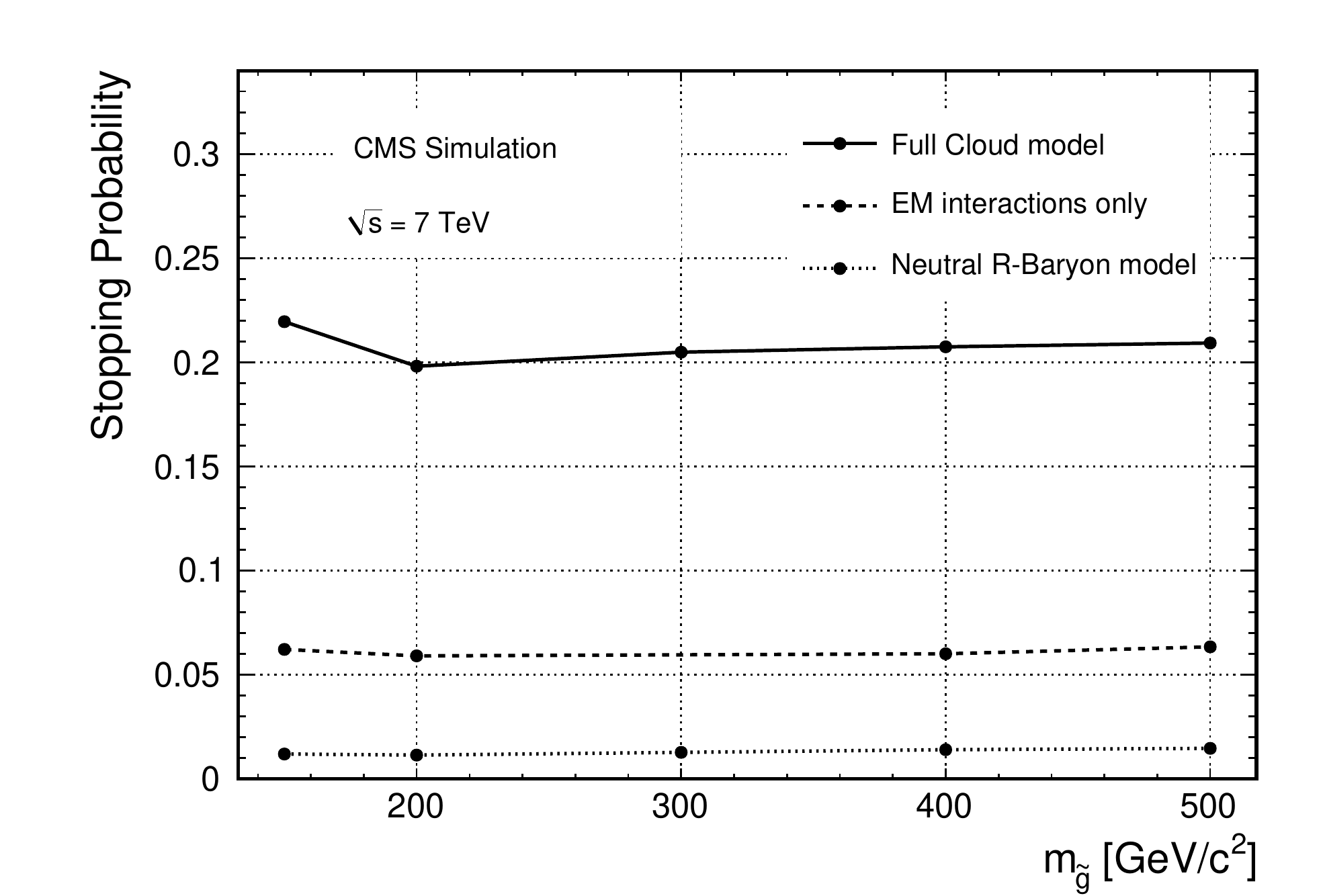}} 
     \caption{Probability for a produced R-hadron to stop anywhere inside the CMS detector
       for different gluino masses and $\sqrt{s} = 7$ TeV.  The solid line shows the stopping probability for the cloud model with both electromagnetic (EM) and nuclear interactions (NI), the dashed line is for the EM only model, and the dotted line is for the ``neutral R-baryon'' model in which only R-mesons stop~\cite{Buccella:1985cs}.  Charge exchange reactions are considered NI.\label{fig:stopEff}}
   \end{center}
 \end{figure}

We employed a dedicated trigger to search for decays of particles at times when there are no collisions.  Information from the beam position and timing (BPTX) monitors are used to identify gaps between the proton bunches that comprise the LHC beam.  The BPTX monitors are positioned 175 m from the center of CMS on either side of the CMS interaction region and produce a signal when an LHC proton bunch passes the monitor. 
Even though the R-hadron decay does not produce a true jet, the resultant energy deposition is sufficiently jet-like that a jet trigger is reasonably efficient. We therefore require a jet trigger together with the condition that a coincidence of signals from both BPTX did not occur, ensuring that the trigger will not fire on jets produced from $pp$ collisions.  The L1 trigger requires a jet with at least 10 GeV transverse energy. A 20 GeV threshold on jet energy is applied in the HLT.  At both L1 and HLT, the pseudorapidity of the jet, $|\eta_{{\rm jet}}|$, is required to be less than 3.0.

Additional selection criteria are applied during data analysis.  Despite the BPTX veto in the trigger, several beam-related processes remain possible sources of background.  In order to reject background events due to an unpaired proton bunch passing through CMS, events in which {\it either} BPTX is over threshold are vetoed. Instrumental effects during trigger generation, and features of the LHC beam such as lower intensity ``satellite'' bunches that accompany the colliding protons, can cause triggers in some of the 25 ns intervals (BX's) which precede or follow the one in which the intended proton collisions occur. We therefore reject any event occurring up to two BX's before, or one BX after, the BX in which collisions are expected. To reject beam-halo muon events, which may not be synchronous with proton collisions, we employ a loose beam-halo veto using the cathode strip chambers (CSC) in the endcap muon system.  The algorithm rejects events in which a beam-halo trigger was recorded, or a track segment was reconstructed in the CSC system with timing consistent with beam-halo, or a muon track was reconstructed with beam-halo-like kinematics.  Finally, to ensure that no out-of-time $pp$ collision events due to satellite bunches contaminate the search sample, events with one or more reconstructed primary vertices are rejected.

A small fraction of cosmic rays traversing the CMS detector deposit significant energy in the calorimeters.  To reduce this background, events that contain reconstructed muons are vetoed. Once beam-related backgrounds and cosmic rays are removed, the remaining source of background is instrumental noise.  Standard calorimeter cleaning and noise rejection criteria~\cite{Chatrchyan:2009hy,HCALPAS,METPAS} are applied.  
We restrict our search to jets in the central HCAL; we require that the most energetic jet in the event has $ |\eta_{{\rm jet}}| < 1.3$.  To suppress noise fluctuations and energy deposits from cosmic rays, a jet with reconstructed energy above 50 GeV is required.  To remove events where a single HCAL channel has misfired, events with more than 90\% of the energy deposited in three or fewer calorimeter towers are vetoed.  We also require that the leading jet has at least 60\% of its energy contained in fewer than 6 towers.  To suppress noise from hybrid photodiode discharge~\cite{Chatrchyan:2009hy}, events with 5 or more of the leading towers at the same azimuthal angle, or where more than 95\% of the jet energy is contained within towers at the same azimuthal angle, are rejected.  

The HCAL electronics have a well-defined time response to charge deposits generated by showering particles. Analog signal pulses produced by these electronics are sampled at 40~MHz, in synchrony with the LHC clock.  These pulses are readout over ten BX samples centered around the pulse maximum.  A physical pulse has some notable properties which we use to distinguish it from noise pulses.  There is a clear peak in the signal pulse shape ($BX_{{\rm peak}}$), significant energy in one bunch crossing before the peak ($BX_{{\rm peak}-1}$), and an exponential decay for several BX's following the peak. We use the ratios $R_1$ = $BX_{{\rm peak}+1} / BX_{{\rm peak}}$ and $R_2$ =  $BX_{{\rm peak}+2}/BX_{{\rm peak}+1}$ to characterize the exponential decay, requiring $R_1 > 0.15$ and $0.10 < R_2 < 0.50$.  Since a physical pulse spans only four time samples we are able to reject noise events based on the presence of energy in previous or successive BX's. We remove events with more than 10\% of the energy of the pulse outside of the central four BX's.  Energy deposits from physical particles tend to have a large fraction of the pulse energy in the peak BX.  Noise does not produce this pulse shape; noise pulses tend to be spread across many BX's or localized in one BX.  We require the ratio of the peak energy to the total energy to be between 0.4 and 0.7.  The requirement that the pulse shape satisfy all the preceding criteria rejects 50\% of the remaining events in the background sample while preserving 93\% of simulated signal events.

We have developed a custom, factorized simulation of gluino production, stopping, and
decay to investigate the experimental signature of this atypical signal.  First, we generate $q\bar{q}\rightarrow\tilde{g}\tilde{g}$
 and $gg\rightarrow\tilde{g}\tilde{g}$ events at $\sqrt{s} = 7$ TeV using {\sc Pythia}~\cite{Sjostrand:2006za}.  The lifetime of the gluino is set such that it is stable.  Gluino masses $m_{\tilde{g}} = 150$ to 500 GeV/$c^2$ are studied.  {\sc Pythia} hadronizes the produced gluino into R-hadrons. A modified {\sc Geant4}~\cite{Agostinelli:2002hh} that implements a ``cloud model'' of heavy stable colored interactions with matter~\cite{Mackeprang:2006gx} is used to simulate the interaction of these R-hadrons with the CMS detector and to record the location at which those R-hadrons that do not exit the detector come to rest.  Figure~\ref{fig:stopEff} presents the stopping probability as a function of gluino mass obtained from our simulation for this model and two alternative models of R-hadronic interactions with matter. Next, we again use {\sc Pythia} to produce an R-hadron at rest which we translate from the nominal vertex position to the recorded stopping location and decay the constituent gluino instantaneously via $\tilde{g} \rightarrow g\,\tilde{\chi}^0_1$.  Finally, we use a specialized Monte Carlo simulation to determine how often the stopped gluino decay would occur during a triggerable beam gap.  Further details of this simulation are described elsewhere~\cite{PAS09}.  

The efficiency with which triggered events pass all selection criteria is estimated from the simulation to be 54\% for a representative gluino decay signal ($m_{\tilde{g}}=300$~ GeV/$c^2$ and $m_{\tilde{\chi}^0_1}=200$~GeV/$c^2$).  This point was chosen to be above existing limits and within the reach of CMS.  The equivalent efficiency with respect to all stopped particles is 17\% since a significant number of R-hadrons stop in uninstrumented regions of the CMS detector where their subsequent decay would not be observable.  For any new physics model that predicts events with sufficient visible energy, $m_{\tilde{g}} - m_{\tilde{\chi}^0_1} > 100$~GeV/$c^2$, this efficiency does not change significantly.  
We measure the background rate in the control sample after all but one of the selection criteria are applied, $R^{N-1}_{{\rm control}}$. We also measure the background rate in the control sample after all selection criteria are applied, $R^{N}_{{\rm control}}$. To obtain an estimate of the background rate in the search sample after all selection criteria, we again measure the rate after omitting one selection criterion and multiply it by the ratio of the rates obtained from the control sample, $R^{N}_{{\rm search}} = R^{N-1}_{{\rm search}} (R^{N}_{{\rm control}}/R^{N-1}_{{\rm control}})$.  This procedure is performed twice, each time omitting one of the most powerful background rejection criteria such that $R^{N-1} \gg R^{N}$; we take the mean of both determinations as the final background rate estimate.

We estimate the systematic uncertainty on the background to be 23\% from the observed variation of the control sample rate $R^{N-1}_{{\rm control}}$ during the time period in which the data were taken. There is also a potential systematic uncertainty due to the accuracy with which the energy deposition of our jet-like signal is simulated. From proton/pion test-beam data and studies of the energy deposited in HCAL by incident cosmic rays, we estimate this introduces a 7\% uncertainty on the acceptance. The systematic uncertainty due to trigger efficiency is negligible since the data analyzed are well above the turn-on region.  Similarly, the systematic uncertainty due to reconstruction efficiency is negligible since we restrict our search to $m_{\tilde{g}} - m_{\tilde{\chi}^0_1} > 100$ GeV/$c^2$ wherein we are fully efficient.  Finally, there is an 11\% uncertainty on the luminosity measurement~\cite{LUMIPAS}.  Limits on a particular model (e.g., gluinos in split supersymmetry) introduce more substantial systematic uncertainties, since the signal yield is sensitive to the stopping probability.  The stopping probability varies greatly depending on the model of R-hadronic interactions used in the simulation.

After the selection criteria described in the preceding paragraphs are applied, we perform a counting experiment and a time-profile analysis on the remaining data.  For the counting experiment, we consider gluino lifetime hypotheses from 75~ns to $10^6$ seconds, where we have chosen the upper limit of the search to be the longest lifetime for which we can still expect to observe at least one event.  For lifetime hypotheses shorter than one LHC orbit (89 $\mu$s),  we search within a time window following each filled bunch crossing.  This time window is equal to $1.256\times \tau_{\tilde{g}}$ for optimal sensitivity to each hypothesized gluino lifetime $\tau_{\tilde{g}}$.  In addition to the lifetimes required to map the general features of the exclusion limit, we include two lifetimes for each observed event:  the largest lifetime hypothesis for which the event lies outside the time window, and the smallest lifetime hypothesis for which the event is contained within the time window.  For lifetime hypotheses longer than one LHC fill, we do not consider the possibility that any observed events may have come from gluinos produced in a previous fill.

In the search sample, we do not observe a significant excess above expected background for any lifetime hypothesis.  The results of this counting experiment for selected lifetime hypotheses are presented in Table~\ref{tab:results}.  In the absence of any discernible signal, we proceed to set 95\%~confidence level (C.L.) limits over 13 orders of magnitude in gluino lifetime using a hybrid CL$_{{\rm S}}$ method~\cite{cls} inspired by Ref.{~\cite{Cousins:1991qz}. 
In Fig.~\ref{fig:limit2syst} we show the 95\%~C.L. limit on $\sigma(pp \rightarrow \tilde{g}\tilde{g})  \times BR(\tilde{g} \rightarrow g\tilde{\chi}^0_1) $ for a mass difference $m_{\tilde{g}}-m_{\tilde{\chi}^0_1} >100$ GeV/$c^2$.  The error bands include statistical and systematic uncertainties.  With the horizontal line in Fig.~\ref{fig:limit2syst} we show a recent NLO+NLL calculation of the cross section at $\sqrt{s} = 7$ TeV for $m_{\tilde{g}} = 300$ GeV/$c^2$ from the authors of Ref.~\cite{Beenakker:2009ha}. 
To illustrate the effect of the stopping probability uncertainty, we present three different 95\%~C.L. limits on $\sigma(pp \rightarrow \tilde{g}\tilde{g})  \times BR(\tilde{g} \rightarrow g\tilde{\chi}^0_1) $ in which the three different R-hadron models are used.  Assuming the cloud model for the interaction of R-hadrons with matter, and assuming BR($\tilde{g} \rightarrow g\tilde{\chi}^0_1$) = 100\%, we are able to exclude lifetimes from 75 ns to $3 \times 10^{5}$ s for $m_{\tilde{g}} = 300$ GeV/$c^2$ with the counting experiment.  Finally, we present the result as a function of the gluino mass in Fig.~\ref{fig:limitMass}. Under the same assumptions as for the cross section limit, we exclude $m_{\tilde{g}} < 370$ GeV/$c^2$ for lifetimes between 10~$\mu$s and 1000~s.  If we assume the EM only model for R-hadronic interactions with matter in order to compare with what was done in Ref.~\cite{Abazov:2007ht}, this exclusion becomes $m_{\tilde{g}} < 302$ GeV/$c^2$.

\begin{table}[th]
\caption{Results of counting experiments for selected values of $\tau_{\tilde{g}}$.  Entries between $1 \times 10^{-5}$ and $1 \times 10^{6}$~s are identical and are suppressed from the table. \label{tab:results}}
\begin{center} 
\begin{tabular}{|l|c|c|}
\hline 
Lifetime [s] & Expected Background ($\pm$ stat. $\pm$ syst.) & Observed \\ \hline
$1 \times 10^{-7}$ & $ 0.8 \pm 0.2 \pm 0.2$ & 2\\
$1 \times 10^{-6}$ & $ 1.9 \pm 0.4 \pm 0.5$ & 3\\
$1 \times 10^{-5}$ & $ 4.9 \pm 1.0 \pm 1.3$ & 5\\
$1 \times 10^{6}$ & $ 4.9 \pm 1.0 \pm 1.3$ & 5\\
\hline 
\end{tabular}
\end{center} 
\end{table}

\begin{figure}[ht]
\begin{center}
  \resizebox{1.0\linewidth}{0.7\linewidth}{\includegraphics{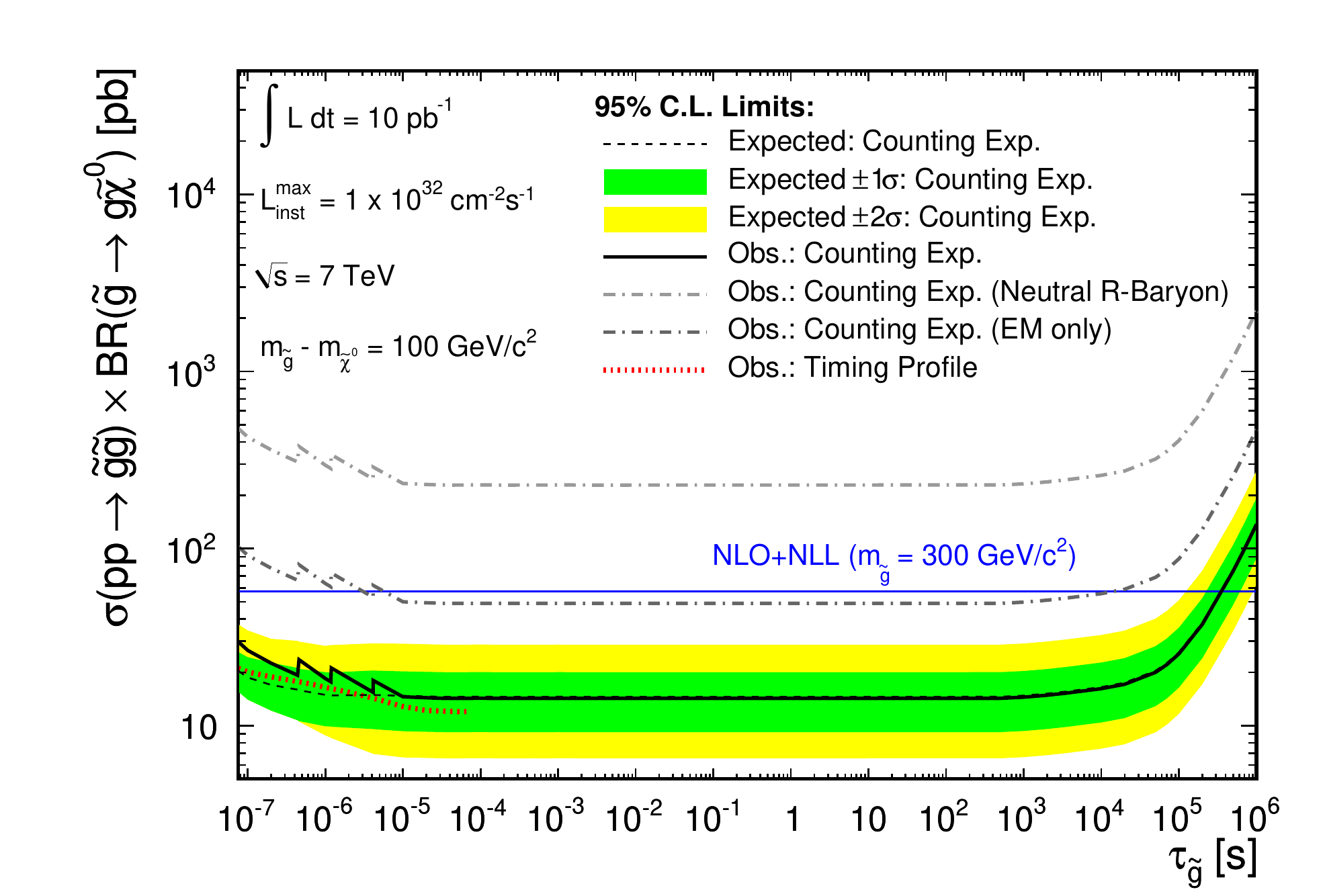}}
  \caption{Expected and observed 95\% C.L. limits on gluino pair production cross section times branching fraction using the ``cloud model'' of R-hadron interactions as a function of gluino lifetime from both the counting experiment and the time-profile analysis. Observed 95\% C.L. limits on the gluino cross section for alternative R-hadron interaction models are also presented.  The NLO+NLL calculation is for $m_{\tilde{g}} = 300$ GeV/$c^2$ from a private communication with the authors of Ref.~\cite{Beenakker:2009ha}.  
\label{fig:limit2syst}}
\end{center}
\end{figure}

\begin{figure}[ht]
\begin{center}
  \resizebox{1.0\linewidth}{0.7\linewidth}{\includegraphics{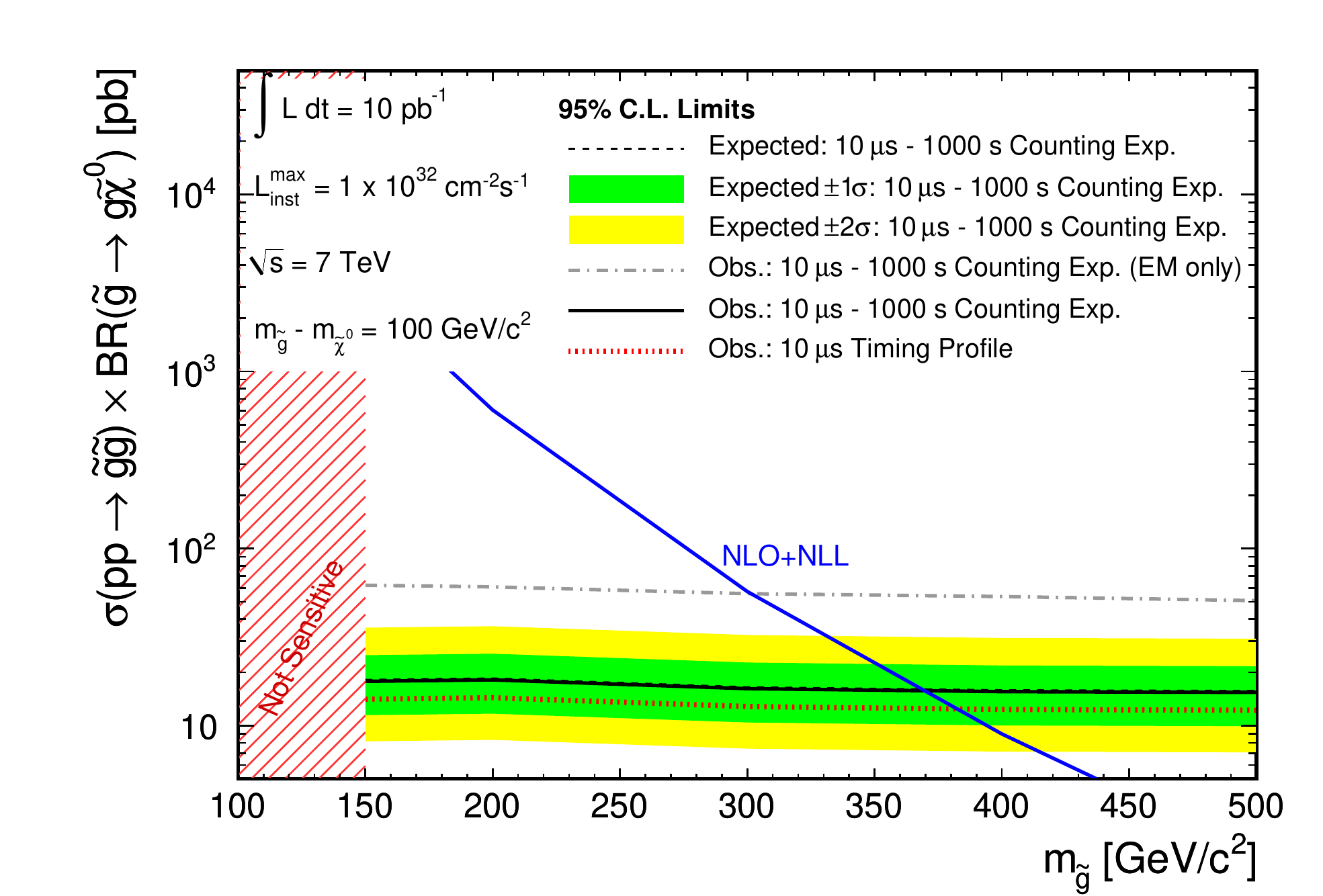}}
  \caption{95\% C.L. limits on gluino pair production cross section times branching fraction as a function of gluino mass assuming the ``cloud model'' of R-hadron interactions (solid line) and EM interactions only (dot-dashed line).  The $m_{\tilde{g}}-m_{\tilde{\chi}^0_1}$ mass difference is maintained at 100 GeV/$c^2$; results are only presented for $m_{\tilde{\chi}^0_1} > 50$ GeV/$c^2$.  The NLO+NLL calculation is from a private communication with the authors of Ref.~\cite{Beenakker:2009ha}.  
The lifetimes chosen are those for which the counting experiment and time-profile analysis are most sensitive
\label{fig:limitMass}}
\end{center}
\end{figure}

We also perform a time-profile analysis.  Whereas, for short lifetimes, a signal from a stopped gluino decay is correlated in time with the collisions, backgrounds are flat in time.  Since the signal and background have very different time profiles, it is possible to extract both their contributions by analyzing the distribution of the observed events in time.
We assume all colliding bunches in an orbit have equal individual instantaneous luminosity.  
We build a probability density function (PDF) for the gluino decay signal as a function of time for a given gluino lifetime hypothesis and the actual times of LHC beam crossings as recorded in our data.  Figure~\ref{fig:inorbitPDFsignal} shows an example of such a PDF for a gluino lifetime of 1~$\mu$s; the in-orbit positions of 2 observed events in the subset of our data that were recorded during an LHC fill with 140 colliding bunches are overlaid.
We limit the range of lifetime hypotheses considered for this time-profile analysis to 75~ns to 100~$\mu$s such that the gluino lifetime is not much longer than the orbit period.  For each lifetime hypothesis we build a corresponding
signal time profile, fit the signal plus background contribution to the data, and extract a 95\%~C.L. upper limit on the possible signal contribution.  The obtained results are plotted as a dotted line in Fig.~\ref{fig:limit2syst}.  This temporal analysis relies only on the flatness of the background shape; it does not have the counting experiment's systematic uncertainty on the background normalization.  Consequently, its dominant systematic uncertainty is the 11\% uncertainty on the luminosity measurement.  For a mass difference $m_{\tilde{g}}-m_{\tilde{\chi}^0_1} >100$ GeV/$c^2$, assuming BR($\tilde{g} \rightarrow g\tilde{\chi}^0_1$) = 100\%, we are able to exclude  $m_{\tilde{g}} < 382$ GeV/$c^2$ at the 95\%~C.L. for a lifetime of 10~$\mu$s with the time-profile analysis.

 \begin{figure}[ht]
 \begin{center}
   \resizebox{1.0\linewidth}{0.5\linewidth}{\includegraphics{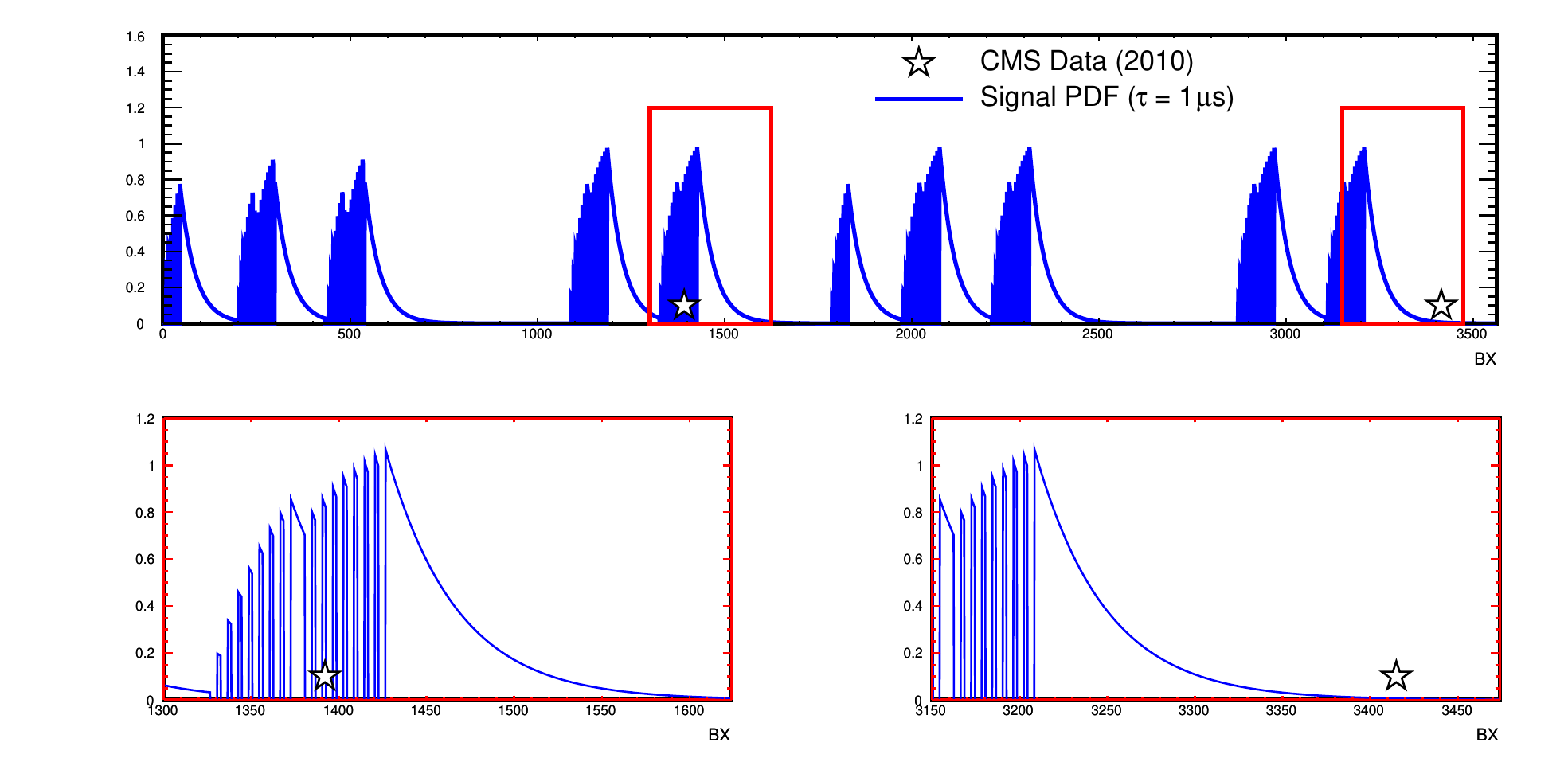}}
   \caption{The top panel shows the in-orbit positions of 2 observed events in the subset of our data that was recorded during an LHC fill with 140 colliding bunches. The decay profile for a 1 $\mu$s lifetime hypothesis is overlaid.  The bottom panels are zoomed views of the boxed regions around the 2 events in the top panel so that the exponential decay shape of the signal hypothesis can be seen. \label{fig:inorbitPDFsignal}}
 \end{center}
 \end{figure}

We have presented the results of the first search for long-lived gluinos produced in 7 TeV $pp$ collisions at the LHC.  We looked for the subsequent decay of those gluinos that would have stopped in the CMS detector during time intervals where there were no $pp$ collisions.  In particular, we searched for decays during gaps in the LHC beam structure.  We recorded such decays with dedicated calorimeter triggers.  In a dataset with a peak instantaneous luminosity of $1 \times 10^{32}~ {\rm cm}^{-2} {\rm s}^{-1}$, an integrated luminosity of 10 pb$^{-1}$, and a search interval corresponding to 62 hours of LHC operation, no significant excess above background was observed.  Limits at the 95\%~C.L. on gluino pair production over 13 orders of magnitude of gluino lifetime are set. For a mass difference $m_{\tilde{g}}-m_{\tilde{\chi}^0_1} >100$ GeV/$c^2$, assuming BR($\tilde{g} \rightarrow g\tilde{\chi}^0_1$) = 100\%, we exclude $m_{\tilde{g}} < 370$ GeV/$c^2$ for lifetimes from 10~$\mu$s to 1000~s with a counting experiment.  Under the same assumptions, we are able to further exclude $m_{\tilde{g}} < 382$ GeV/$c^2$ at the 95\%~C.L. for a lifetime of 10~$\mu$s with a time-profile analysis. These results extend existing limits from the D\O\ Collaboration~\cite{Abazov:2007ht} on both gluino lifetime and gluino mass.  These limits are the most restrictive to date.

We wish to congratulate our colleagues in the CERN accelerator departments for the excellent performance of the LHC machine. We thank the technical and administrative staff at CERN and other CMS institutes, and acknowledge support from: FMSR (Austria); FNRS and FWO (Belgium); CNPq, CAPES, FAPERJ, and FAPESP (Brazil); MES (Bulgaria); CERN; CAS, MoST, and NSFC (China); COLCIENCIAS (Colombia); MSES (Croatia); RPF (Cyprus); Academy of Sciences and NICPB (Estonia); Academy of Finland, ME, and HIP (Finland); CEA and CNRS/IN2P3 (France); BMBF, DFG, and HGF (Germany); GSRT (Greece); OTKA and NKTH (Hungary); DAE and DST (India); IPM (Iran); SFI (Ireland); INFN (Italy); NRF and WCU (Korea); LAS (Lithuania); CINVESTAV, CONACYT, SEP, and UASLP-FAI (Mexico); PAEC (Pakistan); SCSR (Poland); FCT (Portugal); JINR (Armenia, Belarus, Georgia, Ukraine, Uzbekistan); MST and MAE (Russia); MSTD (Serbia); MICINN and CPAN (Spain); Swiss Funding Agencies (Switzerland); NSC (Taipei); TUBITAK and TAEK (Turkey); STFC (United Kingdom); DOE and NSF (USA).

\bibliography{auto_generated}   

\cleardoublepage\appendix\section{The CMS Collaboration \label{app:collab}}\begin{sloppypar}\hyphenpenalty=5000\widowpenalty=500\clubpenalty=5000\input{EXO-10-003-authorlist.tex}\end{sloppypar}
\end{document}

%% file: ptdr-definitions.tex
%
%
%

\providecommand {\etal}{\mbox{et al.}\xspace} 
\providecommand {\ie}{\mbox{i.e.}\xspace}     
\providecommand {\eg}{\mbox{e.g.}\xspace}     
\providecommand {\etc}{\mbox{etc.}\xspace}     
\providecommand {\vs}{\mbox{\sl vs.}\xspace}      
\providecommand {\mdash}{\ensuremath{\mathrm{-}}} 

\providecommand {\Lone}{Level-1\xspace} 
\providecommand {\Ltwo}{Level-2\xspace}
\providecommand {\Lthree}{Level-3\xspace}

\providecommand{\ACERMC} {\textsc{AcerMC}\xspace}
\providecommand{\ALPGEN} {{\textsc{alpgen}}\xspace}
\providecommand{\CHARYBDIS} {{\textsc{charybdis}}\xspace}
\providecommand{\CMKIN} {\textsc{cmkin}\xspace}
\providecommand{\CMSIM} {{\textsc{cmsim}}\xspace}
\providecommand{\CMSSW} {{\textsc{cmssw}}\xspace}
\providecommand{\COBRA} {{\textsc{cobra}}\xspace}
\providecommand{\COCOA} {{\textsc{cocoa}}\xspace}
\providecommand{\COMPHEP} {\textsc{CompHEP}\xspace}
\providecommand{\EVTGEN} {{\textsc{evtgen}}\xspace}
\providecommand{\FAMOS} {{\textsc{famos}}\xspace}
\providecommand{\GARCON} {\textsc{garcon}\xspace}
\providecommand{\GARFIELD} {{\textsc{garfield}}\xspace}
\providecommand{\GEANE} {{\textsc{geane}}\xspace}
\providecommand{\GEANTfour} {{\textsc{geant4}}\xspace}
\providecommand{\GEANTthree} {{\textsc{geant3}}\xspace}
\providecommand{\GEANT} {{\textsc{geant}}\xspace}
\providecommand{\HDECAY} {\textsc{hdecay}\xspace}
\providecommand{\HERWIG} {{\textsc{herwig}}\xspace}
\providecommand{\HIGLU} {{\textsc{higlu}}\xspace}
\providecommand{\HIJING} {{\textsc{hijing}}\xspace}
\providecommand{\IGUANA} {\textsc{iguana}\xspace}
\providecommand{\ISAJET} {{\textsc{isajet}}\xspace}
\providecommand{\ISAPYTHIA} {{\textsc{isapythia}}\xspace}
\providecommand{\ISASUGRA} {{\textsc{isasugra}}\xspace}
\providecommand{\ISASUSY} {{\textsc{isasusy}}\xspace}
\providecommand{\ISAWIG} {{\textsc{isawig}}\xspace}
\providecommand{\MADGRAPH} {\textsc{MadGraph}\xspace}
\providecommand{\MCATNLO} {\textsc{mc@nlo}\xspace}
\providecommand{\MCFM} {\textsc{mcfm}\xspace}
\providecommand{\MILLEPEDE} {{\textsc{millepede}}\xspace}
\providecommand{\ORCA} {{\textsc{orca}}\xspace}
\providecommand{\OSCAR} {{\textsc{oscar}}\xspace}
\providecommand{\PHOTOS} {\textsc{photos}\xspace}
\providecommand{\PROSPINO} {\textsc{prospino}\xspace}
\providecommand{\PYTHIA} {{\textsc{pythia}}\xspace}
\providecommand{\SHERPA} {{\textsc{sherpa}}\xspace}
\providecommand{\TAUOLA} {\textsc{tauola}\xspace}
\providecommand{\TOPREX} {\textsc{TopReX}\xspace}
\providecommand{\XDAQ} {{\textsc{xdaq}}\xspace}

\providecommand {\DZERO}{D\O\xspace}     


\providecommand{\de}{\ensuremath{^\circ}}
\providecommand{\ten}[1]{\ensuremath{\times \text{10}^\text{#1}}}
\providecommand{\unit}[1]{\ensuremath{\text{\,#1}}\xspace}
\providecommand{\mum}{\ensuremath{\,\mu\text{m}}\xspace}
\providecommand{\micron}{\ensuremath{\,\mu\text{m}}\xspace}
\providecommand{\cm}{\ensuremath{\,\text{cm}}\xspace}
\providecommand{\mm}{\ensuremath{\,\text{mm}}\xspace}
\providecommand{\mus}{\ensuremath{\,\mu\text{s}}\xspace}
\providecommand{\keV}{\ensuremath{\,\text{ke\hspace{-.08em}V}}\xspace}
\providecommand{\MeV}{\ensuremath{\,\text{Me\hspace{-.08em}V}}\xspace}
\providecommand{\GeV}{\ensuremath{\,\text{Ge\hspace{-.08em}V}}\xspace}
\providecommand{\TeV}{\ensuremath{\,\text{Te\hspace{-.08em}V}}\xspace}
\providecommand{\PeV}{\ensuremath{\,\text{Pe\hspace{-.08em}V}}\xspace}
\providecommand{\keVc}{\ensuremath{{\,\text{ke\hspace{-.08em}V\hspace{-0.16em}/\hspace{-0.08em}}c}}\xspace}
\providecommand{\MeVc}{\ensuremath{{\,\text{Me\hspace{-.08em}V\hspace{-0.16em}/\hspace{-0.08em}}c}}\xspace}
\providecommand{\GeVc}{\ensuremath{{\,\text{Ge\hspace{-.08em}V\hspace{-0.16em}/\hspace{-0.08em}}c}}\xspace}
\providecommand{\TeVc}{\ensuremath{{\,\text{Te\hspace{-.08em}V\hspace{-0.16em}/\hspace{-0.08em}}c}}\xspace}
\providecommand{\keVcc}{\ensuremath{{\,\text{ke\hspace{-.08em}V\hspace{-0.16em}/\hspace{-0.08em}}c^\text{2}}}\xspace}
\providecommand{\MeVcc}{\ensuremath{{\,\text{Me\hspace{-.08em}V\hspace{-0.16em}/\hspace{-0.08em}}c^\text{2}}}\xspace}
\providecommand{\GeVcc}{\ensuremath{{\,\text{Ge\hspace{-.08em}V\hspace{-0.16em}/\hspace{-0.08em}}c^\text{2}}}\xspace}
\providecommand{\TeVcc}{\ensuremath{{\,\text{Te\hspace{-.08em}V\hspace{-0.16em}/\hspace{-0.08em}}c^\text{2}}}\xspace}

\providecommand{\pbinv} {\mbox{\ensuremath{\,\text{pb}^\text{$-$1}}}\xspace}
\providecommand{\fbinv} {\mbox{\ensuremath{\,\text{fb}^\text{$-$1}}}\xspace}
\providecommand{\nbinv} {\mbox{\ensuremath{\,\text{nb}^\text{$-$1}}}\xspace}
\providecommand{\percms}{\ensuremath{\,\text{cm}^\text{$-$2}\,\text{s}^\text{$-$1}}\xspace}
\providecommand{\lumi}{\ensuremath{\mathcal{L}}\xspace}
\providecommand{\Lumi}{\ensuremath{\mathcal{L}}\xspace}
%
\providecommand{\LvLow}  {\ensuremath{\mathcal{L}=\text{10}^\text{32}\,\text{cm}^\text{$-$2}\,\text{s}^\text{$-$1}}\xspace}
\providecommand{\LLow}   {\ensuremath{\mathcal{L}=\text{10}^\text{33}\,\text{cm}^\text{$-$2}\,\text{s}^\text{$-$1}}\xspace}
\providecommand{\lowlumi}{\ensuremath{\mathcal{L}=\text{2}\times \text{10}^\text{33}\,\text{cm}^\text{$-$2}\,\text{s}^\text{$-$1}}\xspace}
\providecommand{\LMed}   {\ensuremath{\mathcal{L}=\text{2}\times \text{10}^\text{33}\,\text{cm}^\text{$-$2}\,\text{s}^\text{$-$1}}\xspace}
\providecommand{\LHigh}  {\ensuremath{\mathcal{L}=\text{10}^\text{34}\,\text{cm}^\text{$-$2}\,\text{s}^\text{$-$1}}\xspace}
\providecommand{\hilumi} {\ensuremath{\mathcal{L}=\text{10}^\text{34}\,\text{cm}^\text{$-$2}\,\text{s}^\text{$-$1}}\xspace}


\providecommand{\zp}{\ensuremath{\mathrm{Z}^\prime}\xspace}


\providecommand{\kt}{\ensuremath{k_{\mathrm{T}}}\xspace}
\providecommand{\BC}{\ensuremath{\mathrm{B_{c}}}\xspace}
\providecommand{\bbarc}{\ensuremath{\mathrm{\overline{b}c}}\xspace}
\providecommand{\bbbar}{\ensuremath{\mathrm{b\overline{b}}}\xspace}
\providecommand{\ccbar}{\ensuremath{\mathrm{c\overline{c}}}\xspace}
\providecommand{\JPsi}{\ensuremath{\mathrm{J}\hspace{-.08em}/\hspace{-.14em}\psi}\xspace}
\providecommand{\bspsiphi}{\ensuremath{\mathrm{B_s} \to \JPsi\, \phi}\xspace}
\providecommand{\AFB}{\ensuremath{A_\text{FB}}\xspace}
\providecommand{\EE}{\ensuremath{\mathrm{e^+e^-}}\xspace}
\providecommand{\MM}{\ensuremath{\mu^+\mu^-}\xspace}
\providecommand{\TT}{\ensuremath{\tau^+\tau^-}\xspace}
\providecommand{\wangle}{\ensuremath{\sin^{2}\theta_{\text{eff}}^\text{lept}(M^2_\mathrm{Z})}\xspace}
\providecommand{\ttbar}{\ensuremath{\mathrm{t\overline{t}}}\xspace}
\providecommand{\stat}{\ensuremath{\,\text{(stat.)}}\xspace}
\providecommand{\syst}{\ensuremath{\,\text{(syst.)}}\xspace}

\providecommand{\HGG}{\ensuremath{\mathrm{H}\to\gamma\gamma}}
\providecommand{\gev}{\GeV}
\providecommand{\GAMJET}{\ensuremath{\gamma + \text{jet}}}
\providecommand{\PPTOJETS}{\ensuremath{\mathrm{pp}\to\text{jets}}}
\providecommand{\PPTOGG}{\ensuremath{\mathrm{pp}\to\gamma\gamma}}
\providecommand{\PPTOGAMJET}{\ensuremath{\mathrm{pp}\to\gamma + \mathrm{jet}}}
\providecommand{\MH}{\ensuremath{M_{\mathrm{H}}}}
\providecommand{\RNINE}{\ensuremath{R_\mathrm{9}}}
\providecommand{\DR}{\ensuremath{\Delta R}}


\providecommand{\PT}{\ensuremath{p_{\mathrm{T}}}\xspace}
\providecommand{\pt}{\ensuremath{p_{\mathrm{T}}}\xspace}
\providecommand{\ET}{\ensuremath{E_{\mathrm{T}}}\xspace}
\providecommand{\HT}{\ensuremath{H_{\mathrm{T}}}\xspace}
\providecommand{\et}{\ensuremath{E_{\mathrm{T}}}\xspace}
\providecommand{\Em}{\ensuremath{E\hspace{-0.6em}/}\xspace}
\providecommand{\Pm}{\ensuremath{p\hspace{-0.5em}/}\xspace}
\providecommand{\PTm}{\ensuremath{{p}_\mathrm{T}\hspace{-1.02em}/}\xspace}
\providecommand{\PTslash}{\ensuremath{{p}_\mathrm{T}\hspace{-1.02em}/}\xspace}
\providecommand{\ETm}{\ensuremath{E_{\mathrm{T}}^{\text{miss}}}\xspace}
\providecommand{\ETslash}{\ensuremath{E_{\mathrm{T}}\hspace{-1.1em}/}\xspace}
\providecommand{\MET}{\ensuremath{E_{\mathrm{T}}^{\text{miss}}}\xspace}
\providecommand{\ETmiss}{\ensuremath{E_{\mathrm{T}}^{\text{miss}}}\xspace}
\providecommand{\VEtmiss}{\ensuremath{{\vec E}_{\mathrm{T}}^{\text{miss}}}\xspace}

\providecommand{\dd}[2]{\ensuremath{\frac{\mathrm{d} #1}{\mathrm{d} #2}}}

%

\providecommand{\ga}{\ensuremath{\gtrsim}}
\providecommand{\la}{\ensuremath{\lesssim}}
\providecommand{\swsq}{\ensuremath{\sin^2\theta_\mathrm{W}}\xspace}
\providecommand{\cwsq}{\ensuremath{\cos^2\theta_\mathrm{W}}\xspace}
\providecommand{\tanb}{\ensuremath{\tan\beta}\xspace}
\providecommand{\tanbsq}{\ensuremath{\tan^{2}\beta}\xspace}
\providecommand{\sidb}{\ensuremath{\sin 2\beta}\xspace}
\providecommand{\alpS}{\ensuremath{\alpha_S}\xspace}
\providecommand{\alpt}{\ensuremath{\tilde{\alpha}}\xspace}

\providecommand{\QL}{\ensuremath{\mathrm{Q}_\mathrm{L}}\xspace}
\providecommand{\sQ}{\ensuremath{\tilde{\mathrm{Q}}}\xspace}
\providecommand{\sQL}{\ensuremath{\tilde{\mathrm{Q}}_\mathrm{L}}\xspace}
\providecommand{\ULC}{\ensuremath{\mathrm{U}_\mathrm{L}^\mathrm{C}}\xspace}
\providecommand{\sUC}{\ensuremath{\tilde{\mathrm{U}}^\mathrm{C}}\xspace}
\providecommand{\sULC}{\ensuremath{\tilde{\mathrm{U}}_\mathrm{L}^\mathrm{C}}\xspace}
\providecommand{\DLC}{\ensuremath{\mathrm{D}_\mathrm{L}^\mathrm{C}}\xspace}
\providecommand{\sDC}{\ensuremath{\tilde{\mathrm{D}}^\mathrm{C}}\xspace}
\providecommand{\sDLC}{\ensuremath{\tilde{\mathrm{D}}_\mathrm{L}^\mathrm{C}}\xspace}
\providecommand{\LL}{\ensuremath{\mathrm{L}_\mathrm{L}}\xspace}
\providecommand{\sL}{\ensuremath{\tilde{\mathrm{L}}}\xspace}
\providecommand{\sLL}{\ensuremath{\tilde{\mathrm{L}}_\mathrm{L}}\xspace}
\providecommand{\ELC}{\ensuremath{\mathrm{E}_\mathrm{L}^\mathrm{C}}\xspace}
\providecommand{\sEC}{\ensuremath{\tilde{\mathrm{E}}^\mathrm{C}}\xspace}
\providecommand{\sELC}{\ensuremath{\tilde{\mathrm{E}}_\mathrm{L}^\mathrm{C}}\xspace}
\providecommand{\sEL}{\ensuremath{\tilde{\mathrm{E}}_\mathrm{L}}\xspace}
\providecommand{\sER}{\ensuremath{\tilde{\mathrm{E}}_\mathrm{R}}\xspace}
\providecommand{\sFer}{\ensuremath{\tilde{\mathrm{f}}}\xspace}
\providecommand{\sQua}{\ensuremath{\tilde{\mathrm{q}}}\xspace}
\providecommand{\sUp}{\ensuremath{\tilde{\mathrm{u}}}\xspace}
\providecommand{\suL}{\ensuremath{\tilde{\mathrm{u}}_\mathrm{L}}\xspace}
\providecommand{\suR}{\ensuremath{\tilde{\mathrm{u}}_\mathrm{R}}\xspace}
\providecommand{\sDw}{\ensuremath{\tilde{\mathrm{d}}}\xspace}
\providecommand{\sdL}{\ensuremath{\tilde{\mathrm{d}}_\mathrm{L}}\xspace}
\providecommand{\sdR}{\ensuremath{\tilde{\mathrm{d}}_\mathrm{R}}\xspace}
\providecommand{\sTop}{\ensuremath{\tilde{\mathrm{t}}}\xspace}
\providecommand{\stL}{\ensuremath{\tilde{\mathrm{t}}_\mathrm{L}}\xspace}
\providecommand{\stR}{\ensuremath{\tilde{\mathrm{t}}_\mathrm{R}}\xspace}
\providecommand{\stone}{\ensuremath{\tilde{\mathrm{t}}_1}\xspace}
\providecommand{\sttwo}{\ensuremath{\tilde{\mathrm{t}}_2}\xspace}
\providecommand{\sBot}{\ensuremath{\tilde{\mathrm{b}}}\xspace}
\providecommand{\sbL}{\ensuremath{\tilde{\mathrm{b}}_\mathrm{L}}\xspace}
\providecommand{\sbR}{\ensuremath{\tilde{\mathrm{b}}_\mathrm{R}}\xspace}
\providecommand{\sbone}{\ensuremath{\tilde{\mathrm{b}}_1}\xspace}
\providecommand{\sbtwo}{\ensuremath{\tilde{\mathrm{b}}_2}\xspace}
\providecommand{\sLep}{\ensuremath{\tilde{\mathrm{l}}}\xspace}
\providecommand{\sLepC}{\ensuremath{\tilde{\mathrm{l}}^\mathrm{C}}\xspace}
\providecommand{\sEl}{\ensuremath{\tilde{\mathrm{e}}}\xspace}
\providecommand{\sElC}{\ensuremath{\tilde{\mathrm{e}}^\mathrm{C}}\xspace}
\providecommand{\seL}{\ensuremath{\tilde{\mathrm{e}}_\mathrm{L}}\xspace}
\providecommand{\seR}{\ensuremath{\tilde{\mathrm{e}}_\mathrm{R}}\xspace}
\providecommand{\snL}{\ensuremath{\tilde{\nu}_L}\xspace}
\providecommand{\sMu}{\ensuremath{\tilde{\mu}}\xspace}
\providecommand{\sNu}{\ensuremath{\tilde{\nu}}\xspace}
\providecommand{\sTau}{\ensuremath{\tilde{\tau}}\xspace}
\providecommand{\Glu}{\ensuremath{\mathrm{g}}\xspace}
\providecommand{\sGlu}{\ensuremath{\tilde{\mathrm{g}}}\xspace}
\providecommand{\Wpm}{\ensuremath{\mathrm{W}^{\pm}}\xspace}
\providecommand{\sWpm}{\ensuremath{\tilde{\mathrm{W}}^{\pm}}\xspace}
\providecommand{\Wz}{\ensuremath{\mathrm{W}^{0}}\xspace}
\providecommand{\sWz}{\ensuremath{\tilde{\mathrm{W}}^{0}}\xspace}
\providecommand{\sWino}{\ensuremath{\tilde{\mathrm{W}}}\xspace}
\providecommand{\Bz}{\ensuremath{\mathrm{B}^{0}}\xspace}
\providecommand{\sBz}{\ensuremath{\tilde{\mathrm{B}}^{0}}\xspace}
\providecommand{\sBino}{\ensuremath{\tilde{\mathrm{B}}}\xspace}
\providecommand{\Zz}{\ensuremath{\mathrm{Z}^{0}}\xspace}
\providecommand{\sZino}{\ensuremath{\tilde{\mathrm{Z}}^{0}}\xspace}
\providecommand{\sGam}{\ensuremath{\tilde{\gamma}}\xspace}
\providecommand{\chiz}{\ensuremath{\tilde{\chi}^{0}}\xspace}
\providecommand{\chip}{\ensuremath{\tilde{\chi}^{+}}\xspace}
\providecommand{\chim}{\ensuremath{\tilde{\chi}^{-}}\xspace}
\providecommand{\chipm}{\ensuremath{\tilde{\chi}^{\pm}}\xspace}
\providecommand{\Hone}{\ensuremath{\mathrm{H}_\mathrm{d}}\xspace}
\providecommand{\sHone}{\ensuremath{\tilde{\mathrm{H}}_\mathrm{d}}\xspace}
\providecommand{\Htwo}{\ensuremath{\mathrm{H}_\mathrm{u}}\xspace}
\providecommand{\sHtwo}{\ensuremath{\tilde{\mathrm{H}}_\mathrm{u}}\xspace}
\providecommand{\sHig}{\ensuremath{\tilde{\mathrm{H}}}\xspace}
\providecommand{\sHa}{\ensuremath{\tilde{\mathrm{H}}_\mathrm{a}}\xspace}
\providecommand{\sHb}{\ensuremath{\tilde{\mathrm{H}}_\mathrm{b}}\xspace}
\providecommand{\sHpm}{\ensuremath{\tilde{\mathrm{H}}^{\pm}}\xspace}
\providecommand{\hz}{\ensuremath{\mathrm{h}^{0}}\xspace}
\providecommand{\Hz}{\ensuremath{\mathrm{H}^{0}}\xspace}
\providecommand{\Az}{\ensuremath{\mathrm{A}^{0}}\xspace}
\providecommand{\Hpm}{\ensuremath{\mathrm{H}^{\pm}}\xspace}
\providecommand{\sGra}{\ensuremath{\tilde{\mathrm{G}}}\xspace}
\providecommand{\mtil}{\ensuremath{\tilde{m}}\xspace}
\providecommand{\rpv}{\ensuremath{\rlap{\kern.2em/}R}\xspace}
\providecommand{\LLE}{\ensuremath{LL\bar{E}}\xspace}
\providecommand{\LQD}{\ensuremath{LQ\bar{D}}\xspace}
\providecommand{\UDD}{\ensuremath{\overline{UDD}}\xspace}
\providecommand{\Lam}{\ensuremath{\lambda}\xspace}
\providecommand{\Lamp}{\ensuremath{\lambda'}\xspace}
\providecommand{\Lampp}{\ensuremath{\lambda''}\xspace}
\providecommand{\spinbd}[2]{\ensuremath{\bar{#1}_{\dot{#2}}}\xspace}

\providecommand{\MD}{\ensuremath{{M_\mathrm{D}}}\xspace}
\providecommand{\Mpl}{\ensuremath{{M_\mathrm{Pl}}}\xspace}
\providecommand{\Rinv} {\ensuremath{{R}^{-1}}\xspace} 

%% file: EXO-10-003-authorlist.tex
\textbf{Yerevan Physics Institute,  Yerevan,  Armenia}\\*[0pt]
V.~Khachatryan, A.M.~Sirunyan, A.~Tumasyan
\vskip\cmsinstskip
\textbf{Institut f\"{u}r Hochenergiephysik der OeAW,  Wien,  Austria}\\*[0pt]
W.~Adam, T.~Bergauer, M.~Dragicevic, J.~Er\"{o}, C.~Fabjan, M.~Friedl, R.~Fr\"{u}hwirth, V.M.~Ghete, J.~Hammer\cmsAuthorMark{1}, S.~H\"{a}nsel, C.~Hartl, M.~Hoch, N.~H\"{o}rmann, J.~Hrubec, M.~Jeitler, G.~Kasieczka, W.~Kiesenhofer, M.~Krammer, D.~Liko, I.~Mikulec, M.~Pernicka, H.~Rohringer, R.~Sch\"{o}fbeck, J.~Strauss, A.~Taurok, F.~Teischinger, W.~Waltenberger, G.~Walzel, E.~Widl, C.-E.~Wulz
% ...
\vskip\cmsinstskip
% ...
1:~~Also at CERN, European Organization for Nuclear Research, Geneva, Switzerland\\
% ...
